\documentclass[prl,twocolumn,aps,unsortedaddress,showpacs,letterpaper]{revtex4-1}
\usepackage{graphicx}
\usepackage{epsfig}
\usepackage{natbib}

\begin{document}

\title{Ubiquity of Linear Resistivity at Intermediate Temperature in Bad Metals}

\author{G.~R.~Boyd}
\affiliation{Department of Physics, Georgetown University, Washington, DC 20057, USA}
\author{V.~Zlati\'c}
\affiliation{Institute of Physics, Zagreb POB 304, Croatia}
\affiliation{Department of Physics, Georgetown University, Washington, DC 20057, USA}
\author{J.~K.~Freericks}
\affiliation{Department of Physics, Georgetown University, Washington, DC 20057, USA}

\begin{abstract}
Bad metals display transport behavior that differs from what is commonly seen in ordinary metals. One of the most significant differences is a resistivity that is linear in temperature and rises to well above the Ioffe-Regel limit (where the mean-free path is equal to the lattice spacing). Using an exact Kubo formula, we show that a linear resistivity naturally occurs for many systems when they are in an incoherent intermediate-temperature state. We verify the analytic arguments with numerical calculations for a simplified version of the Hubbard model which is solved with dynamical mean-field theory. Similar features have also been seen in Hubbard models, where they can begin at even lower temperatures due to the formation of resilient quasiparticles.
\end{abstract}

\date{\today}
\maketitle

Transport properties of strongly correlated materials, such as oxides in the families of
vanadates ~\cite{urano.2000}, cobaltates~\cite{kriener_2004} or cuprates~\cite{hussey_2004},  
Kondo semiconductors such as FeSi ~\cite{Manyala_2008,sales.2011}, FeSb$_2$ ~\cite{jie_2012} 
CeB$_6$\cite{Kuni_2000} or SmB$_6$\cite{Cooley_1995}, 
and organic charge transfer salts~\cite{Organics} 
are poorly understood, despite an overwhelming amount of experimental work 
which established non-Fermi-liquid behavior for these systems~\cite{StewartRMP,ImadaRMP}. 
In particular, a resistivity which rises linearly with temperature above the Mott-Ioffe-Regel 
limit~\cite{ioffe-regel} has become a hallmark for non-Fermi liquid behavior~\cite{RevModPhys.79.1015}. 
One common feature of these vastly different materials is that they are formed by doping away from 
a Mott-Hubbard insulating state. Starting from this observation, and the ubiquity of quasi-linear non-Fermi 
liquid materials, we provide a simple explanation of the experimental data.

We begin by deriving the transport coefficients using an analytic approach, in the spirit of Mahan and Sofo's work 
on the best thermoelectrics~\cite{MahanSofo}, where the 
optimization of transport properties was calculated based on a simplified ansatz for the transport relaxation 
time which then allowed one to perform the optimization.  
Here, we work in a similar vein, but consider the temperature dependence of the resistivity based on a general 
discussion of the properties of the transport relaxation time for a strongly correlated metal. 
By modeling this simplest form for correlated transport, the results should hold for a wide range of materials, and thereby explain the ubiquity 
of the linear resistivity at intermediate temperature. 
In the second part, we substantiate the phenomenological results by calculating 
the resistivity of a non-trivial model of strongly correlated electrons propagating 
on a $d$-dimensional lattice. We use the Falicov-Kimball model which, 
like the Hubbard or periodic Anderson model, has a gap in the excitation spectrum 
and, unlike these other models,  admits an exact solution for the resistivity at arbitrary 
doping and temperature. 

Our starting point is the Kubo formula for the conductivity which reads~\cite{mahan.81}, 
\begin{equation}
\sigma_{dc}(T)=\sigma_{0}\sum_\sigma \int d \omega
\left(-\frac{df(\omega)}{d\omega}\right) \tau_\sigma(\omega)~,
\label{eq: kubo}
\end{equation}
where $\sigma_{0}$ is a material specific constant with units of conductivity, 
$\left(-{df(\omega)}/{d\omega}\right)$ is the derivative of the Fermi function that 
is sharply peaked around the chemical potential $\mu$, so that the integral is cut-off outside 
the Fermi window $|\omega| \geq  k_B T$.
The summation is over the spin states $\sigma$ and  $\tau_{\sigma}(\omega)$ is the exact transport relaxation time 
which  includes the velocity factors, averaged over the Fermi surface, and the effects of vertex corrections, if present. 
We set $k_B=\hbar=1$ and measure all energies with respect to $\mu$. 

\begin{figure}[thb!]
\centering
\vskip-0.32in
\includegraphics[width=0.99\columnwidth,clip]{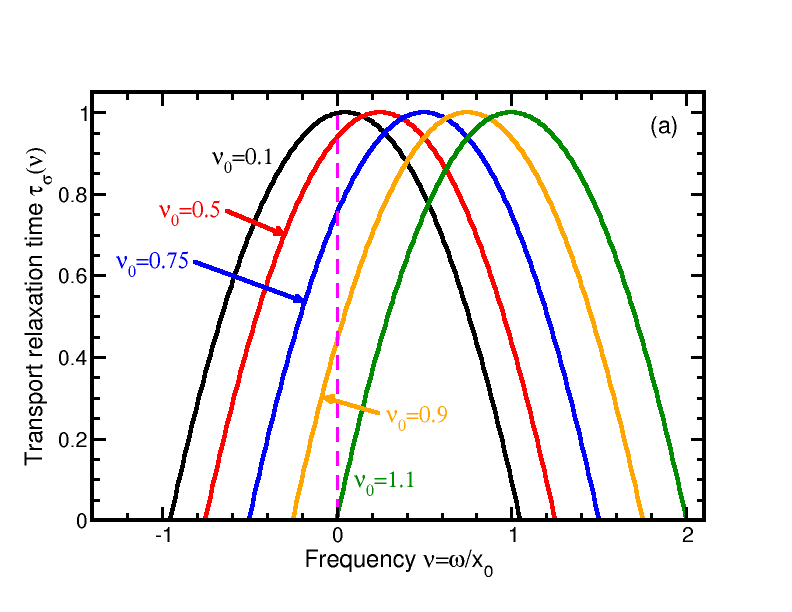}
\vskip-0.14in
\includegraphics[width=0.99\columnwidth,clip]{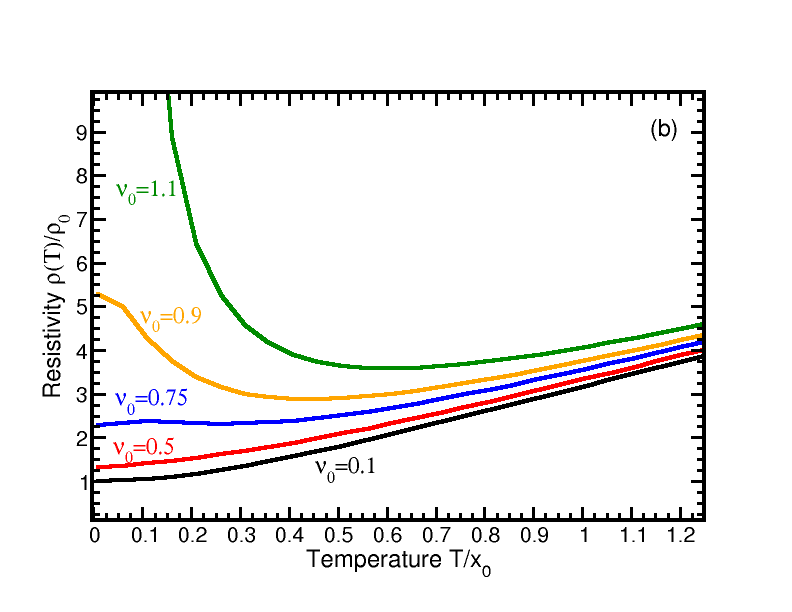}
\vskip-0.18in
\caption[]{(color online) 
Panel (a): Rescaled relaxation time $\tilde\tau_{\sigma}(\nu)$ plotted as
a function of rescaled frequency $\nu=\omega/x_0$ relative to the chemical potential,
$\mu$, which is
indicated by the vertical line at $\nu=0$. (For definition of the scaling factors see the text.)
The different curves show $\tau_{\sigma}(\nu)$ shifted with respect to $\mu$ by $\nu_0=$ 0.1, 0.5, 0.75, 0.9, and 1.1, respectively. 
Curve (a) corresponds to a dirty metal, curves (b), (c), and (d) to a bad metal, and curve  (e) to a slightly doped Mott insulator. 
Panel (b): The rescaled resistivity obtained from Eq.~(\ref{eq: fermi}) plotted as a function of rescaled temperature $\tilde T=T/x_0$. 
The different curves are obtained for $\tilde\tau_{\sigma}(\nu)$ as defined in the panel (a). 
}
\label{figs}
\end{figure}

Since $\tau_{\sigma}(\omega)$  is nonnegative and vanishes for energies outside the band, 
it must have at least one maximum within the band. 
In a Fermi liquid,  $\tau_{\sigma}(\omega)$ diverges as $T\to 0$ and $\omega\to 0$, 
and the resistivity, $\rho(T)=1/ \sigma_{dc}(T)$, follows a $T^2$ law at low temperature. 
If there is residual scattering, due to disorder for example, 
the divergence gets cut-off and the Fermi-liquid form no longer holds. 
In a pure strongly correlated metal, for temperatures {\it above} the low-temperature coherence scale,  
 the transport relaxation time typically has two maxima, located in the upper and the lower Hubbard bands, and
neither the shape nor the position of these broad maxima, relative to $\mu(T)$, change appreciably with temperature. 
The transport relaxation time of the Hubbard model, Falicov-Kimball model, Anderson model, and 
other effective models of strong correlations, exhibits these features. 
Since the chemical potential of a strongly correlated metal is within one of the two Hubbard bands, 
we calculate the resistivity focusing on $\tau_{\sigma}(\omega)$ with just a single broad maximum at $\omega_0$, 
neglecting  the excitations across the gap.  

The conductivity given by Eq.~(\ref{eq: kubo}) crucially depends on the overlap between $(-df/d\omega)$  
and $\tau_{\sigma}(\omega)$, i.e., on temperature and doping. 
Temperature broadens the Fermi window where the integrand is appreciable, while doping changes the number of carriers, 
so that $\mu$ gets shifted with respect to $\omega_0$.  The value and the shape of $\tau_{\sigma}(\omega)$ 
around $\omega_0$ can also be doping dependent. 

To estimate the resistivity we expand $\tau_{\sigma}(\omega)$ around its maximum at $\omega_0$, 
\begin{equation}
\tau_{\sigma}(\omega)\approx
\tau_0-\frac{\tau_1}{2}(\omega-\omega_0)^2~,
\label{eq: tau_max}
\end{equation}
where $\tau_0 =\tau_{\sigma}(\omega_0)$, $\tau_1=-{d^2 \tau_{\sigma}(\omega)}/{d \omega^2}\Big |_{\omega\rightarrow\omega_0}$, 
and we use a simple model in which $\tau_{\sigma}(\omega)$ is approximated by the parabolic form in Eq.~(\ref{eq: tau_max})
for $ \Lambda_{-} < \omega <\Lambda_{+}$ and $\tau_{\sigma}(\omega)=0$ otherwise; this form properly has a maximum, and shows linear behavior as one approaches the band edges, as expected for a three-dimensional material.
The cutoffs $\Lambda_{\pm}$ are obtained by setting  $ \tau_{\sigma}(\omega)=0$ in Eq.~(\ref{eq: tau_max}). 
This yields $\Lambda_\pm=\omega_0\pm x_0$, where 
$x_0^2={\tau_0/\tau_1}$ is inversely proportional to the curvature of $\tau_{\sigma}(\omega)$ at $\omega_0$ and $x_0$ has dimensions of energy.  
Since the high-energy part of $\tau_{\sigma}(\omega)$ does not contribute much to the conductivity, 
$x_0=\omega_0- \Lambda_{-}$ often defines an effective bandwidth relevant for transport of a doped Mott insulator. 

To evaluate the integral in Eq.~(\ref{eq: kubo}), we introduce dimensionless variables, 
$\nu=\omega/x_0$ and $\tilde T=T/x_0$, and write the relaxation time as,  
$\tau_{\sigma}(\nu)/\tau_0=1-(\nu-\nu_0)^2$, where $\nu_0=\omega_0/x_0$. 
Integrating  by parts, and using $\tau_\sigma(\Lambda_{-})  =\tau_\sigma(\Lambda_{+})=0$, yields 
\begin{equation}
\sigma_{dc}(\tilde T)=
2 \tau_0\sigma_0
\int_{\nu_0-1}^{\nu_0+1} d \nu ~
f(\nu) ~ \frac{d \tau(\nu)}{d\nu} ~, 
\label{eq: fermi}
\end{equation}
where $f(\nu)=1/[1+\exp(\nu/\tilde T)]$, $ d{\tau_a}/{d\nu}=2(\nu-\nu_0)$, and  
we took the spin degeneracy into account. 
The integrand is a regular function and the numerical evaluation is straightforward. 
The renormalized resistivity, $\rho(\tilde T)/\rho_0$, where $\rho_0=1/(\sigma_0\tau_0)$, 
is shown in panel (b) of Fig.~\ref{figs} for several characteristic values of $\nu_0$. 
Panel (a) shows  $\tau_{\sigma}(\nu)/\tau_0$ used for each of the resistivity curves.  
The data indicate three types of behavior, depending on the relative position of $\mu$  and $\omega_0$. Here $\mu$ is fixed, but as seen below, fixing the density produces similar results.

\begin{figure}[thb!]
\centering
\includegraphics[angle=0,width=0.95\columnwidth]{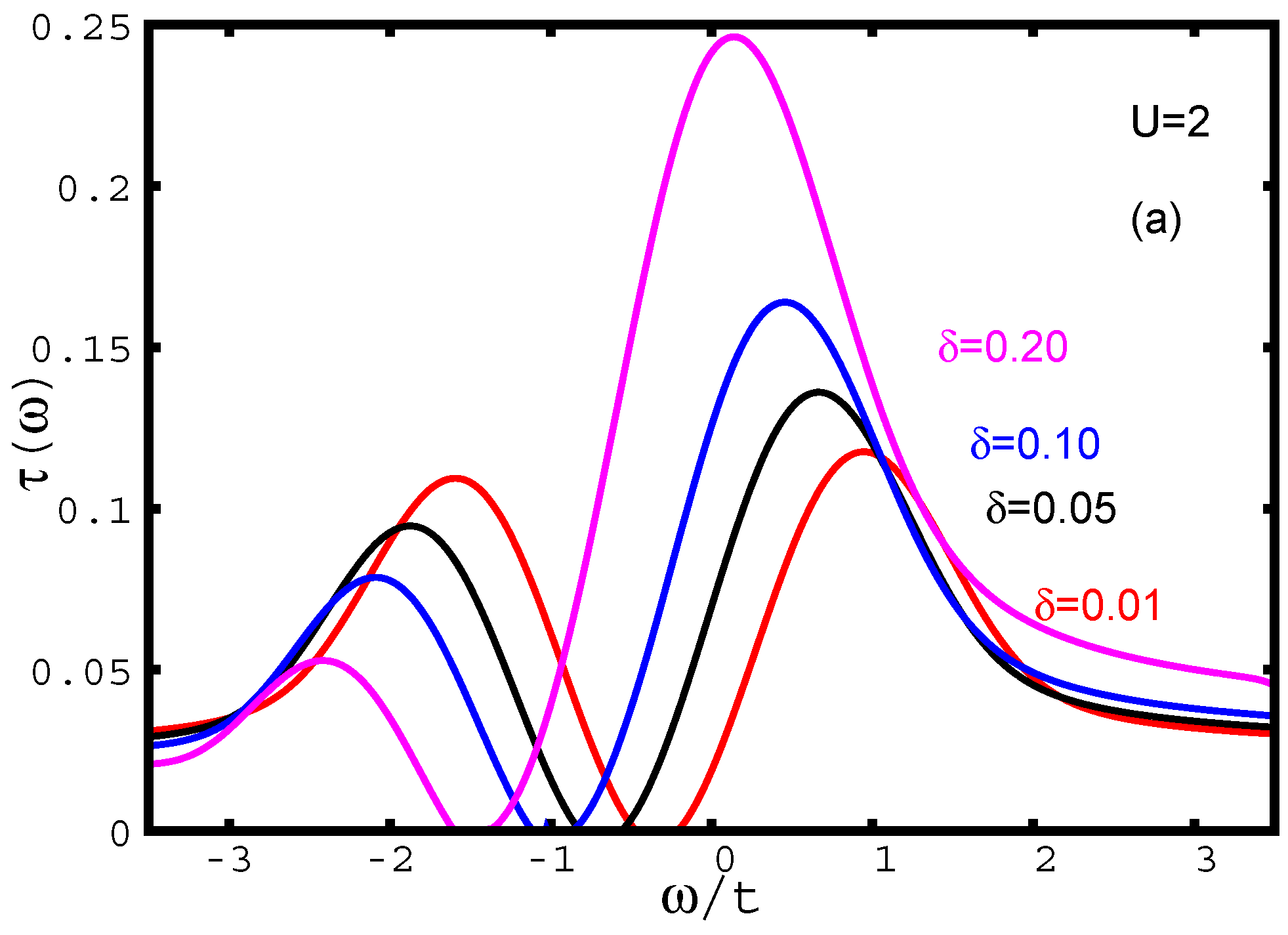} 
\includegraphics[angle=0,width=0.95\columnwidth]{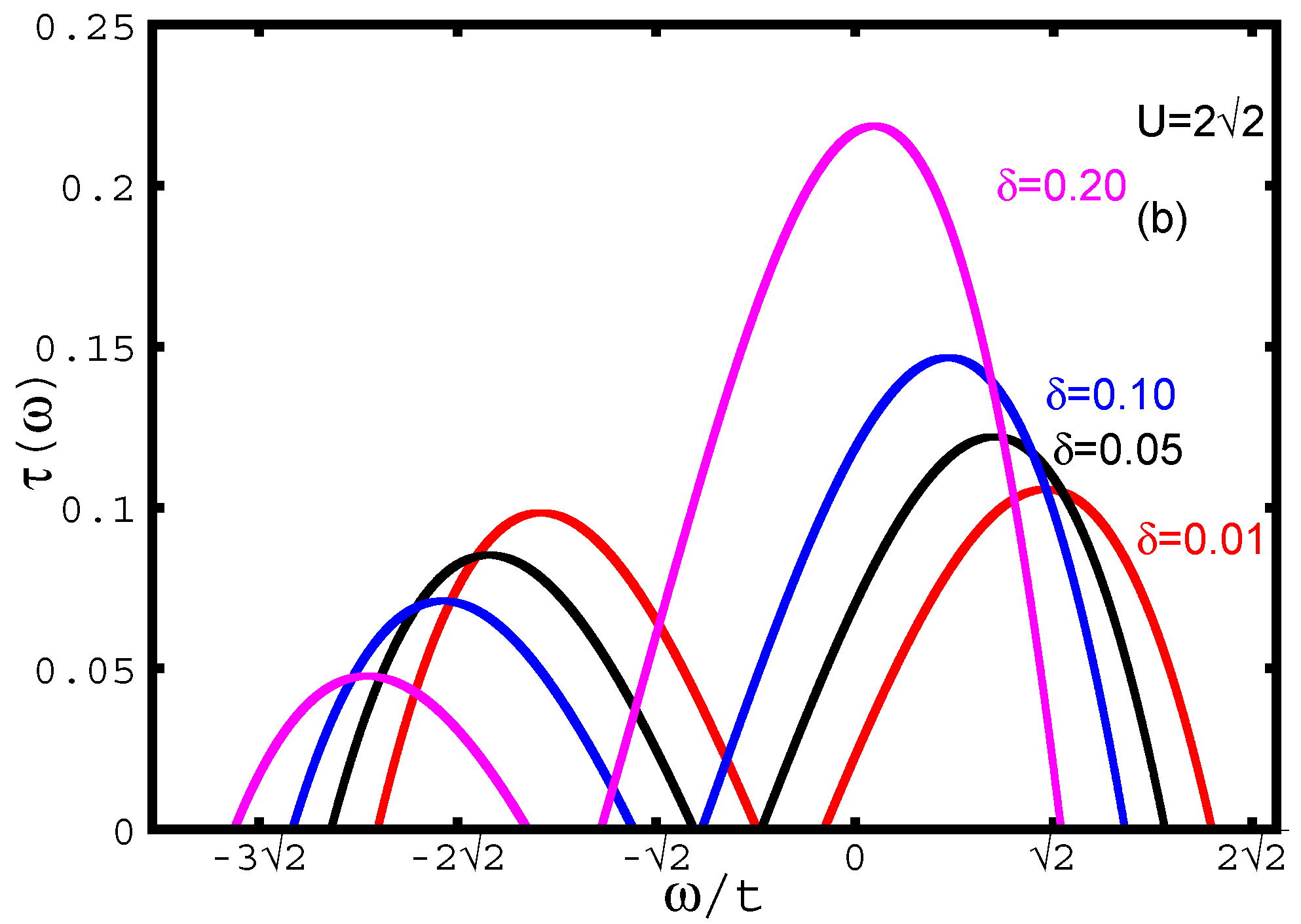}
\caption[]{(color online) Transport lifetimes in the Falicov-Kimball model for $U=2t^*$ on a hypercubic
lattice (a) and $U=2\sqrt{2}t^*$ on the Bethe lattice (b).}  
\label{ScatteringFig}
\end{figure}

For $\nu_0 \geq 1$, when the chemical potential is close to the band-edge, 
the resistivity decreases rapidly as temperature increases from $T=0$.
At about $T\simeq \omega_0/2$, the resistivity drops to a minimum and, 
then, increases with temperature, assuming at about $T\simeq \omega_0$ a linear form. 
Such a behavior is typical of lightly doped Mott insulators. 
For $\nu_0\leq 1$, when the chemical potential is just above the band edge, 
the low-temperature resistivity is metallic. 
It starts from a finite value, at $T=0$, and grows to a well pronounced maximum, which 
is reduced and shifted to lower temperature as $\nu_0$ is reduced. 
The minimum still occurs at about $T\simeq \omega_0/2$ and, for $T\geq \omega_0$, 
the resistivity becomes a linear function in a broad temperature range. Such a behavior is typical of bad metals. 
For $\nu_0 \ll 1$, the chemical potential is close to the maximum of $\tau_{\sigma}(\nu)$ 
and $\rho(T)$ increases parabolically from its zero-temperature value, as found in dirty metals. 
At higher temperatures, $T > \omega_0$, there is a crossover to the linear behavior. 
According to this simple model, strongly correlated materials are classified into three distinct groups: 
lightly doped insulators characterized by a low-temperature resistivity upturn, 
bad metals characterized by an extended range of quasilinear resistivity, 
and dirty metals characterized by a {\em constant plus $T^2$} behavior.  

The analytic approach is suggestive of the robustness of the linear resistivity for bad metals due to the general nature of $\tau_{\sigma}(\omega)$, 
but we want to go further to  obtain similar results with a nontrivial microscopic model. We choose the spin-1/2 Falicov-Kimball 
model which is closely related to the  Hubbard model and leads to similar transport properties (above the 
coherence temperature of the Hubbard model). The question we primarily want to address is: to what extent can a model for strongly 
correlated electrons capture the phenomenology of non-Fermi liquid electrical 
transport with a focus on the linear resistivity? The advantage of the Falicov-Kimball  model is that the dynamical 
mean-field theory (DMFT) provides an exact solution at arbitrary filling~\cite{RevModPhys.75.1333}. 
(There have been related studies on the Hubbard model using DMFT \cite{PhysRevLett.111.036401,PhysRevB.61.7996} exploring transport in bad metals as well). 

The spin-1/2 Falicov-Kimball Hamiltonian  reads
\begin{equation}
H=-\frac{t^*}{2\sqrt{d}}\sum_{\langle i,j\rangle\sigma}c^\dagger_{i\sigma}\ c_{j\sigma}+
U\sum_{i\sigma}w_i 
c^{\dagger}_{i\sigma}c_{i\sigma},
\label{eq: ham}
\end{equation}
where $c^{\dagger}_{i\sigma}\; (c_{i\sigma})$ is the mobile electron creation (annihilation) operator of spin $\sigma$ and 
$w_i$ is 1 or 0 and represents the localized electron number operator at site $i$. 
(Each lattice site can only be occupied by a single localized electron, because the on-site  repulsion between the  
localized electrons of the opposite spin is assumed infinite.)
The interaction of the conduction electrons with localized electrons is $U$ 
and $t^*$ is the hopping integral scaled so that we can properly take the $d \rightarrow \infty$ limit~\cite{PhysRevLett.62.324}. 
We work on both a hypercubic and Bethe lattice using units where $t^*=1$. 
We maintain the paramagnetic constraint,  $\rho_{c\sigma}=\rho_{c\bar\sigma}=\rho_c$, 
by equating the conduction and localized densities. 
For hole doping, we have $\rho_c=\rho_f=1-\delta \leq 1$, where 
$\delta$ is the concentration of the holes in the lower Hubbard band, while
for electron doping, $\rho_c=1+\delta\geq 1$, where $\delta$ is the concentration of electrons in the upper Hubbard band.

The model is solved using DMFT \cite{brandt_mielsch} in the infinite dimensional limit $d \rightarrow \infty$, 
such that the self-energy $\Sigma(\omega)$ is a functional of the local conduction electron Green's function, $G_{loc}(\omega)$, 
and the full lattice problem is equivalent to a single-site  model with an electron coupled self-consistently 
to a time-dependent external field.
Several reviews, whose notation we adopt, now exist both on DMFT generally \cite{DMFTRMP}
and on the exact DMFT for the Falicov-Kimball model \cite{RevModPhys.75.1333}. 
We find $\Sigma(\omega)$, $G_{loc}(\omega)$, and the local density of conduction states 
$ \rho_{loc}(\omega)=- {\rm Im}~ G_{loc}(\omega)/\pi$ numerically using methods
 described elsewhere \cite{PhysRevLett.69.168} . 

\begin{figure}[thb!]
\centering
\includegraphics[angle=0,width=0.95\columnwidth]{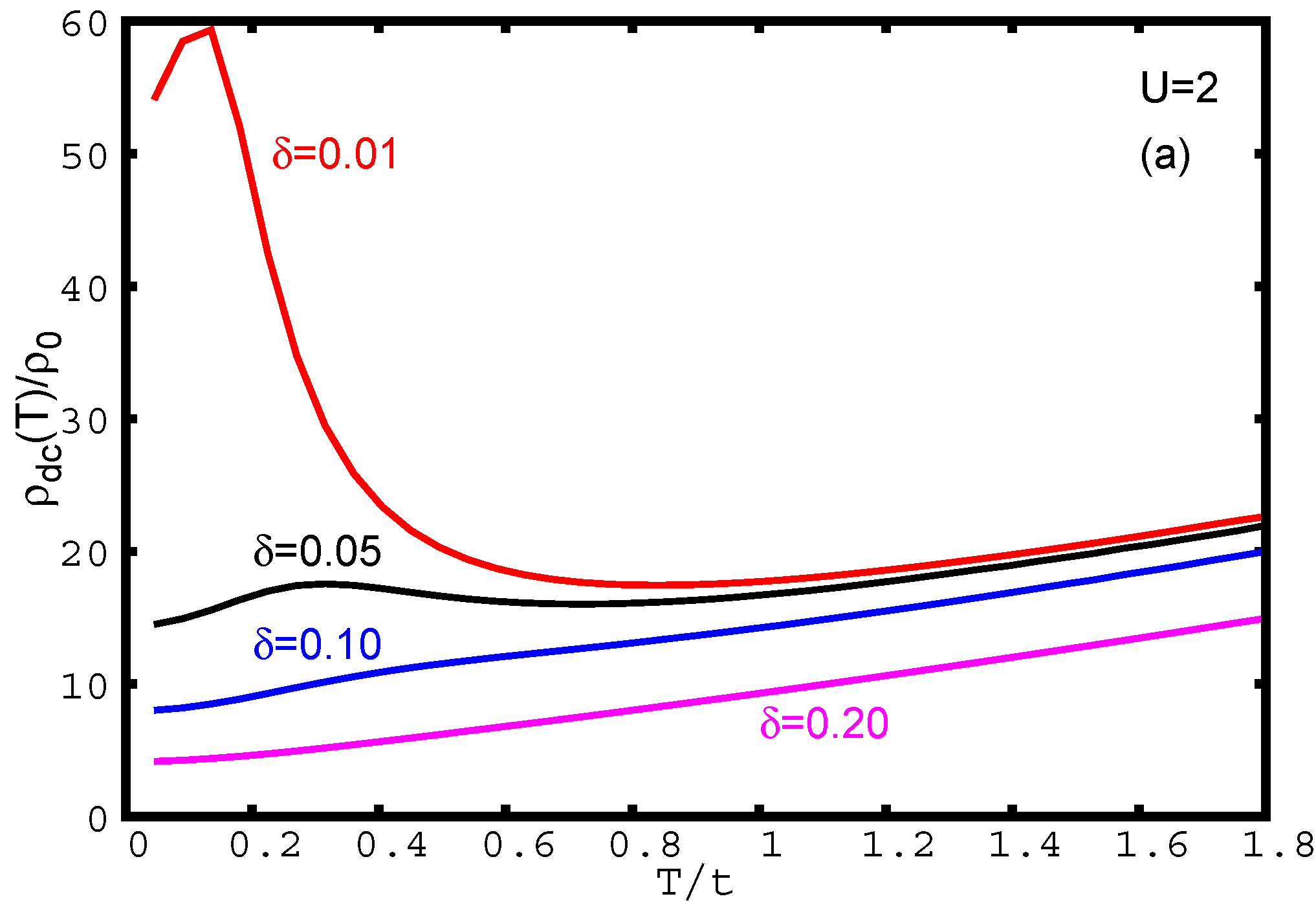} 
\includegraphics[angle=0,width=0.9\columnwidth]{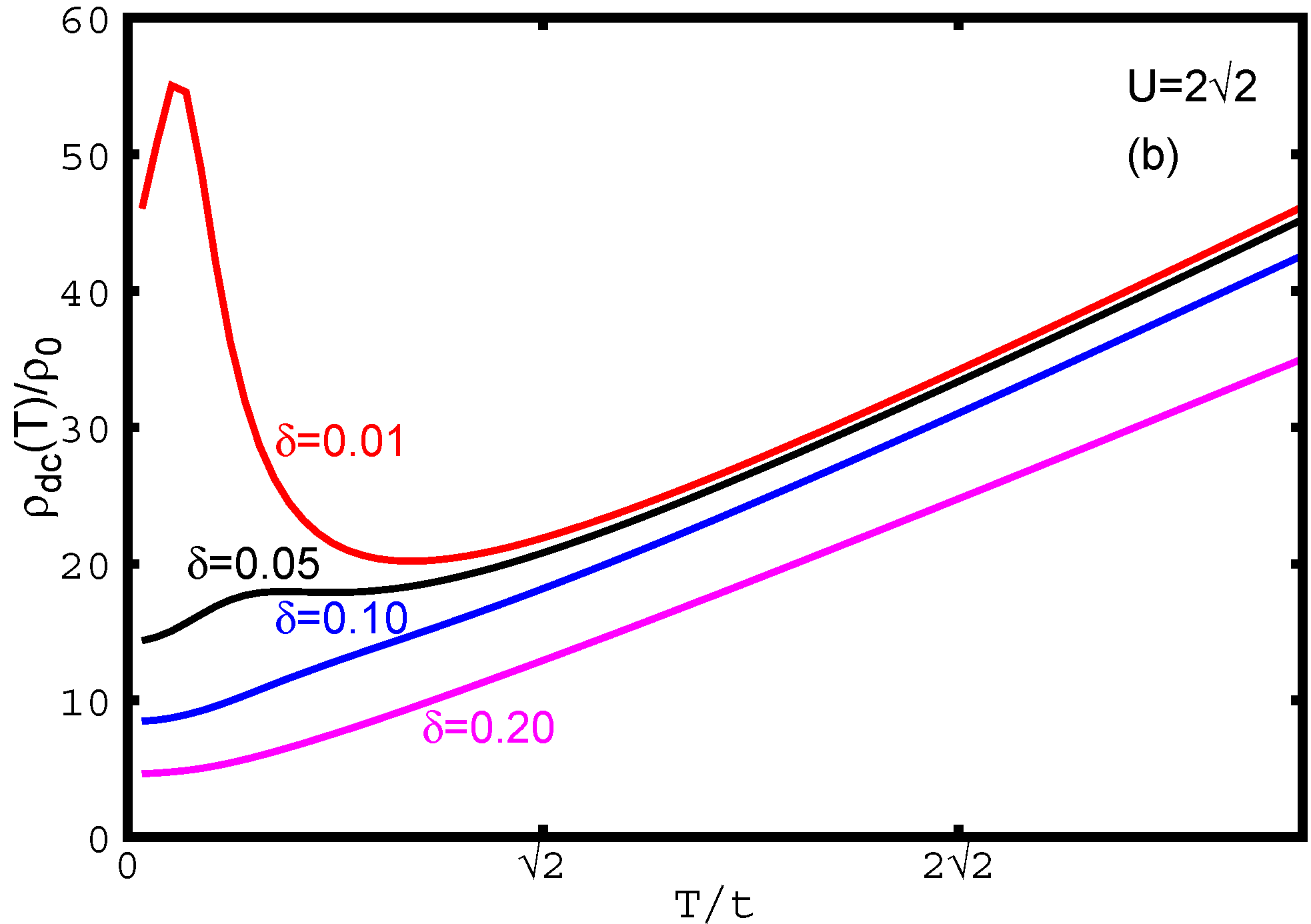}
\caption[]{(color online) Resistivity for (a) $U=2t$ on the hypercubic lattice and (b)
$U=2\sqrt{2}t$ on the Bethe lattice}
\label{rhofig}
\end{figure}

For $\rho_c=1$, $ \rho_{loc}(\omega)$ is symmetric and, for large enough $U$, 
we have a Mott insulator in which a filled lower Hubbard band is separated from an empty upper Hubbard band by a band gap with
the chemical potential in the middle of the gap ($U_c=\sqrt{2}$ for the hypercubic lattice and $U_c=2$ for the Bethe lattice). 
Away from half-filling,  $\rho_{loc}(\omega)$ 
is asymmetric and for electron doping, which is the case we consider, the chemical potential is in the upper Hubbard band.
Its distance from the lower band edge $\Lambda_-$ is determined by charge conservation 
$\delta=2\int d\omega f(\omega) \rho_{loc}(\omega) -1$. 

For $d \rightarrow \infty$, the vertex corrections to the conductivity vanish~\cite{zlatic_horvatic_1990} 
and explicit formulas can be found for the relaxation time.
On the Bethe lattice, this yields~\cite{Ref75}:  
\begin{equation}
\label{Bethe}
\tau_{\sigma}(\omega)=
\frac{1}{3 \pi^2}
\rm{Im}^2[G_{loc}(\omega)] \left (
\frac{|G_{loc}(\omega)|^2-3 }{|G_{loc}(\omega)|^2-1} \right )
~.
\end{equation}
while on the hypercubic lattice,  we have~\cite{RevModPhys.75.1333}:
\begin{eqnarray}
                            \label{hypecubic}
~\tau_{\sigma}(\omega) =&~&
\frac{1}{4 \pi^2}
 \frac{{\rm Im}~ G_{loc}(\omega)}{{\rm Im}~ \Sigma(\omega)} \\
&+& 
 \frac{ 1}{2 \pi^2}
  \left\{ 1-{\rm Re} \left[ \left(\omega+\mu-\Sigma(\omega)\right) G_{loc}(\omega)\right] \right\} ~.
\nonumber
\end{eqnarray}
For fixed $\rho_f$,  the shape of $\tau_{\sigma}(\omega)$ is independent of temperature. 
In a Fermi liquid, where one can approximate~\cite{mahan.81}  
$\tau_{\sigma}(\omega)\simeq {{\rm Im}~ G_{loc}(\omega)}/{{\rm Im}~ \Sigma(\omega)}$ 
with ${\rm Im}\Sigma(\omega\to 0)\to 0$, the relaxation time $\tau_{\sigma}(\omega)$ diverges as $\omega\to 0$.  
In the Falicov-Kimball model,  however, ${\rm Im}~ \Sigma(0)$ does not vanish and $\tau_{\sigma}(0)$ remains finite. 
For large $U$, the width of the  single-particle excitations exceeds their energy leading to  overdamped 
excitations rather than with quasiparticles, such that the Fermi liquid description is not applicable.

The transport relaxation time of  the Falicov-Kimball model due to such overdamped excitations,   
obtained for a fixed value of $U$ and several values of $\delta$, is shown in Fig.~\ref{ScatteringFig}. 
The left and right panel show the results for the hypercubic and Bethe lattice, respectively. Note the similarity to the inverse quadratic approximation used in the first part.
The transport relaxation time vanishes below the band edge $\Lambda_{-}$  and has a peak at the energy $\omega_0$,  
in the upper Hubbard band (for electron doping).  As $\delta$ increases, $\omega_0$ and $\Lambda_{-}$ decrease 
but the difference $\omega_0-\Lambda_{-}$ remains approximately constant. 
The resistivity obtained for the same set of parameters
is shown in Fig.~\ref{rhofig}. 
The doping dependence of $\rho(T)$ follows from the observation that $\delta$ reduces $\omega_0$ and that, 
for $\Lambda_{-} < \mu  < \omega_0$, the Fermi window removes the contribution of the high-energy part of $\tau_{\sigma}(\omega)$. 
Close to half-filling (very small $\delta$), where $\mu\simeq \Lambda_{-}\ll\omega_0$, 
the resistivity exhibits a low-temperature peak, then, drops to a minimum at about $T\simeq \omega_0/2$ and, 
eventually, becomes a linear function of $T$, for $T\geq \omega_0$. 
An increase of $\delta$ brings $\omega_0$ closer to  $\mu$, which reduces the resistivity maximum 
and brings the onset of the linear region to lower temperatures.  
For a sufficiently large $\delta$, the maximum is completely suppressed and the resistivity is a  monotonically increasing  function of temperature. 
For $\delta\simeq 0.2$, we find $\omega_0\simeq\mu$ and obtain a resistivity with a well defined $T^2$ term at the lowest temperatures.
Note, the crossover between different regimes can also be induced by pressure which modifies the hopping integrals 
and shifts  $\omega_0$ with respect to $\mu$. 

The results obtained for the Falicov-Kimball model are in complete agreement with the phenomenological theory 
presented in the first part of the paper.    
Hence, the analytic model is verified as providing the generic behavior of a doped Mott insulator at intermediate $T$.
The central result of this paper is that the linear resistivity seen in strongly correlated
materials at intermediate $T$ is governed by the appearance of a maximum in $\tau_{\sigma}(\omega)$
above the chemical potential. The slope of the linear resistivity does not vary much for a range of chemical potentials near
the maximum, so the temperature dependence of $\mu(T)$ does not change this behavior. 
In other correlated models like the Hubbard model, the linear resistivity will
disappear when $T$ is reduced below the renormalized Fermi-liquid scale, but it appears that
the resilient quasiparticle picture~\cite{PhysRevLett.110.086401} allows the linear region to
be brought down to even lower
$T$'s than seen in the Falicov-Kimball model.

\section{Acknowledgements}

JKF and VZ were supported by the National Science Foundation under grant number DMR-1006605 and the Ministry of Science of Croatia 
for the data analysis. JKF and GRB were supported by the Department of Energy, Office of Basic Energy Sciences, 
under grant number DE-FG02-08ER46542 for the development of the numerical analysis and the development of the analytic model. 
The collaboration was supported the Department of Energy, Office of Basic Energy Sciences, Computational Materials 
and Chemical Sciences Network grant number DE-SC0007091. JKF was also supported by the McDevitt bequest at Georgetown University.

\bibliography{biblio}

\end{document}